# Knowledge Management: An *MIS Quarterly* Research Curation

**Research Curation Team:**

Peng Huang *(University of Maryland)*

Atreyi Kankanhalli *(National University of Singapore)*

Harris Kyriakou *(IESE Business School)*

Rajiv Sabherwal *(University of Arkansas)*

## 1. Focus of the Research Curation

The notions of knowledge and its management have been at the core of the information systems (IS) field almost since its inception. Knowledge has been viewed in several ways in the prior literature, including as a state of mind, an object, a process, access to information, and a capability. A commonly-used definition characterizes knowledge as a justified belief that increases an entity's capacity for effective action (Alavi and Leidner 2001, p. 109). Relatedly, knowledge management (KM) has been defined as a systemic process to acquire, organize, and communicate individual knowledge so that others may make use of it (Beck et al. 2014). Knowledge-management systems (KMSs) support these processes for creating, exchanging, and storing knowledge (Beck et al. 2014), and have been viewed as being either repository-based or network-based (Kankanhalli et al. 2005).

In an attempt to provide a useful resource for scholars interested in KM, we take stock of the pertinent research published in *MISQ*. More specifically, the goal of this curation is to serve as a living document that will offer a starting point for future KM research. This curation highlights the 44 articles with a primary focus on KM (Table 1). The articles address theoretical and conceptual issues, provide methodological guidance, and use a wide range of quantitative and qualitative research methods. To define the scope of this curation, we excluded: (1) articles in which KM is used as part of another construct; (2) some early articles that were practice-oriented with limited scholarly orientation; and (3) articles that focus on knowledge (such as the knowledge requirements of IS professionals) but not on KM.

## 2. Progression of Research in MISQ

The first two research papers in *MISQ* on KM (Meyer and Curley 1991; Byrd et al. 1992) were published in the early 1990's. They focused on knowledge-based systems, or expert systems, and examined the management and development of these systems in organizations (Byrd et al. 1992; Meyer and Curley 1991). Research on KM processes also emerged during this initial period, including a study of the antecedents and outcomes of shared knowledge of IS units in organizations (Nelson and Cooprider 1996). KM research interest in IS started growing following two conceptual papers – one that presented knowledge, KM, and KMS definitions and classifications (Alavi and Leidner 2001), and the other that examined the scientific discourses in IS research on KM (Schultze and Leidner 2002). Other papers during this period advanced the notion that an excessive focus on technology in KM may not be useful (Markus et al. 2002; Massey et al. 2002). In sum, the initial KM publications focused almost exclusively on KM within organizations, and examined KM processes, the design of KM systems, and the antecedents of KM.

KM research gained momentum with the *MISQ* special issue on IT and KM published in 2005, which included twelve papers across two volumes. A third of the papers focused on knowledge sharing, KMS use, and their antecedents (Bock et al. 2005; Kankanhalli et al. 2005; Ko et al. 2005; Wasko and Faraj 2005). Several papers (Malhotra et al. 2005; Van de Ven 2005; Ko et al. 2005; Wasko and Faraj 2005) reflected a transition from the prior exclusive focus on KM within organizations to the consideration of KM beyond organizational boundaries. For example, Malhotra et al. (2005) examined knowledge creation in supply-chain relationships, while Wasko and Faraj (2005) investigated individuals' knowledge contribution in an online network of practice for legal professionals, and Ko et al. uncovered the antecedents of knowledge transfer between external consultants and clients during ERP implementations. Another third of the papers (Chen and Edgington 2005; Lin et al. 2005; Ryu et al. 2005; Tanriverdi 2005) in the special issue were the initial economics-based studies on KM and examined consequences of KM for the firm. Studies during the rest of the decade continued the shift towards examining the impacts of KM, such as in making complex decisions (Arnold et al. 2006) and on IT project performance (Mitchell 2006). They also continued the extension of KM research beyond traditional boundaries, such as in constructing knowledge alliances between multiple KMS for land management in India (Puri 2007) and in managing knowledge in offshore teams (Leonardi and Bailey 2008).

KM publications since 2010 have been increasingly interested in strategies for managing knowledge, in the use of KM systems, and in the consequences of KM. For instance, studies have examined ways to enhance KM, such as through transactive memory systems (Choi et al. 2010; Majchrzak et al. 2013), social media (Leonardi 2015), metamodels (Kyriakou et al. 2017), network structures in crowdsourced communities (Lu et. 2017), and visual ontologies (Bera et al. 2011). In addition, the impacts of KM – such as on innovation (Carlo et al. 2012; Trantopoulos et al. 2017), individuals' careers (Huang and Zhang 2016), job performance (Zhang et al. 2017), team performance (Choi et al. 2010), and firm performance (Iyengar et al. 2015) – have also received considerable attention during this period.

In terms of methodology, we see a diverse range, including conceptual papers (e.g., Alavi et al. 2001; Griffith et al. 2003; Van de Ven 2005); case studies (e.g., Garud and Kumaraswamy 2005; Kotlarsky et al. 2014); laboratory experiments (e.g., Poston and Speier 2005; Bera et al. 2011); field experiments (e.g., Arnold et al. 2006; Leonardi 2015); questionnaire surveys of individuals (Bock et al. 2005; Kankanhalli et al. 2005), matched pairs (e.g., Ko et al. 2005), and teams (Choi et al. 2010); analytical models (e.g., Cha et al. 2008; Ryu et al. 2005); objective data from organizations (e.g., Kim et al. 2016) and enterprise blogging messages (e.g., Beck et al. 2014); econometric modeling of longitudinal data from community networks (e.g., Huang and Zhang 2016); and mixed-methods using qualitative and quantitative methods (e.g., Zhang 2017). Cumulatively, these articles have made significant contributions over time to our understanding of KM strategies, processes, antecedents and outcomes, and the design and use of KM systems, thereby enabling intellectual and practical progress in the field of KM.

### 3. Thematic Advances in Knowledge

Four broad themes emerge from our analysis of the articles in this curation: (1) the strategies for managing knowledge (KM Strategy); (2) the processes for knowledge management (KM Process); (3) the design considerations for KM systems (KMS Design); and (4) the use of KM systems (KMS Use). However, some articles (e.g., Alavi and Leidner 2001; Schultze and Leidner 2002), with a focus on literature review and theoretical development, could not be clearly classified as within one of these four categories. Moreover, the articles within each theme have been conducted with different units of analysis, such as individual/team, organization, or beyond organizational boundaries.

The first theme of KM research we identified focuses on KM strategy. These articles examine various aspects of KM strategy and how they relate to organizational contexts as well as influence performance. As may be expected, most of these studies aim at gaining an understanding of KM at the organizational level. For example, Massey et al. (2002) developed insights regarding a KM strategy that involved elements of process, people, and technology, and explored how this strategy related to both the organization's context and performance. Van de Ven (2005) argued that for organizations seeking to develop and commercialize knowledge-intensive technologies, a strategy of "running in packs" may be more successful than "going it alone."

A second dominant theme of prior KM research has been the processes involved in KM, such as knowledge transfer, knowledge brokering, or knowledge conversion. Many of these studies have been conducted at the individual or team level. For example, Schultze (2000) found that knowledge workers employ three informing practices, i.e., expressing, monitoring, and translating, while striving to balance subjectivity and objectivity in the process of producing information. Nelson and Cooprider (1996) reported that the shared knowledge between IS and line groups mediates the effects of trust and influence on IS performance. More recently, Kyriakou et al. (2017) examined knowledge reuse by a community of product designers and identified reuse for customization as a new process that is distinct from knowledge reuse processes proposed previously, i.e., reuse for replication and reuse for innovation. Additionally, a few articles have pursued this theme at the organizational level. For example, in the context of real-estate franchising, Iyengar et al. (2015) found that IT use is an important learning mechanism for franchisees as it impacts knowledge transfer effectiveness and absorptive capacity, the latter of which affects financial performance.

As a third theme, researchers have investigated issues associated with the design of KMS, and suggested a set of design guidelines and principles. At the individual/team level, Bera et al. (2011) through experiments showed that the use of visual ontology enables users to learn about concepts and relationships relevant to the knowledge domain, and then proposed guidelines for designing visual ontologies for knowledge identification. At the organizational level, Meyer and Curley (1991) emphasized that expert systems usually involve two distinct types of complexity – knowledge complexity and technological complexity – and proposed specific variables to measure each. Finally, several studies have examined KMS design issues crossing organizational boundaries. For example, based on a field case study, Puri (2007) highlighted the importance of constructing knowledge alliances and the need to draw upon a multiplicity of knowledge systems to produce relevant 'hybridized' knowledge for supporting effective IS development and implementation.

The fourth theme reflects a significant research interest in the use of KMS by individuals and teams. For instance, studies have investigated the roles of individual motivations in promoting employees' knowledge contribution to electronic knowledge repositories (Kankanhalli et al. 2005) and the conditions under which repository KMS use leads to superior job performance (Kim et al. 2016; Zhang 2017). Yet others have examined how identification and internalization explain individuals' use of KMS (Wang et al. 2013) and how individuals' use, in terms of contribution to and learning from online communities, affects their job-hopping (or voluntary turnover) behavior (Huang and Zhang 2016). Additionally, for knowledge-based systems, prior research has explored how novice and expert decision makers use explanation facilities differently to make high-level, complex judgments within a cooperative problem-solving environment (Arnold et al. 2006).

## 4. Conclusion

This curation shows the breadth of coverage on the topic of KM, both thematically and methodologically. It also illustrates the rich phenomena and problems in this area. Going forward, we believe that KM and related systems will continue to be important topics of research interest, with knowledge-based economies becoming the engine of growth around the globe. As we see a greater amount of knowledge incorporated into IT (e.g., artificial intelligence systems that are an evolution of earlier knowledge-based systems) and more knowledge being created by IT (e.g., knowledge discovery and data mining from big data), KM topics will continue to draw IS researchers' interest in the future. The contributions made by the articles in this curation provide a solid foundation for future research on this vitally important topic.

**Please cite this curation as follows:** Huang, P., Kankanhalli, A., Kyriakou, H., Sabherwal, R., "Knowledge Management," in *MIS Quarterly* Research Curations, Ashley Bush and Arun Rai, Eds., http://misq.org/research-curations, April 30, 2018.

**Table 1. *MIS Quarterly* Papers on Knowledge Management**

| ID | Author(s) | Title | Year | Vol. | Iss. |
|---|---|---|---|---|---|
| 1 | Marc H. Meyer and Kathleen Foley Curley | An Applied Framework for Classifying the Complexity of Knowledge-Based Systems | 1991 | 15 | 4 |
| 2 | Terry Anthony Byrd, Kathy L. Cossick, and Robert W. Zmud | A Synthesis of Research on Requirements Analysis and Knowledge Acquisition Techniques | 1992 | 16 | 1 |
| 3 | Kay M. Nelson and Jay G. Cooprider | The Contribution of Shared Knowledge to IS Group Performance | 1996 | 20 | 4 |
| 4 | Ulrike Schultze | A Confessional Account of an Ethnography about Knowledge Work | 2000 | 24 | 1 |
| 5 | Maryam Alavi and Dorothy E. Leidner | Review: Knowledge Management and Knowledge Management Systems: Conceptual Foundations and Research Issues | 2001 | 25 | 1 |
| 6 | M. Lynne Markus, Ann Majchrzak, and Les Gasser | A Design Theory for Systems that Support Emergent Knowledge Processes | 2002 | 26 | 3 |
| 7 | Anne P. Massey, Mitzi M. Montoya-Weiss, Tony M. O'Driscoll | Knowledge Management in Pursuit of Performance: Insights from Nortel Networks | 2002 | 26 | 3 |
| 8 | Ulrike Schultze and Dorothy E. Leidner | Studying Knowledge Management in Information Systems Research: Discourses and Theoretical Assumptions | 2002 | 26 | 3 |
| 9 | Terri L. Griffith, John E. Sawyer, and Margaret A. Neale | Virtualness and Knowledge in Teams: Managing the Love Triangle of Organizations, Individuals, and Information Technology | 2003 | 27 | 2 |
| 10 | Suzanne D. Pawlowski and Daniel Robey | Bridging User Organizations: Knowledge Brokering and the Work of Information Technology Professionals | 2004 | 28 | 4 |
| 11 | Gee-Woo Bock, Robert W. Zmud, Young-Gul Kim, and Jae-Nam Lee | Behavioral Intention Formation in Knowledge Sharing: Examining the Roles of Extrinsic Motivators, Social-Psychological Forces, and Organizational Climate | 2005 | 29 | 1 |
| 12 | Raghu Garud and Arun Kumaraswamy | Vicious and Virtuous Circles in the Management of Knowledge: The Case of Infosys Technologies | 2005 | 29 | 1 |
| 13 | Atreyi Kankanhalli, Bernard C.Y. Tan, and Kwok-Kee Wei | Contributing Knowledge to Electronic Knowledge Repositories: An Empirical Investigation | 2005 | 29 | 1 |
| 14 | Dong-Gil Ko, Laurie J., and William R. King | Antecedents of Knowledge Transfer from Consultants to Clients in Enterprise System Implementations | 2005 | 29 | 1 |
| 15 | Arvind Malhotra, Sanjay Gosain, and Omar A. El Sawy | Absorptive Capacity Configurations in Supply Chains: Gearing for Partner-Enabled Market Knowledge Creation | 2005 | 29 | 1 |
| 16 | Molly McLure Wasko and Samer Faraj | Why Should I Share? Examining Social Capital and Knowledge Contribution in Electronic Networks of Practice | 2005 | 29 | 1 |

| 17 | Andrew N.K. Chen and Theresa M. Edgington | Assessing Value in Organizational Knowledge Creation: Considerations for Knowledge Workers | 2005 | 29 | 2 |
| --- | --- | --- | --- | --- | --- |
| 18 | Lihui Lin, Xianjun Geng, and Andrew B. Whinston | A Sender-Receiver Framework for Knowledge Transfer | 2005 | 29 | 2 |
| 19 | Robin S. Poston and Cheri Speier | Effective Use of Knowledge Management Systems: A Process Model of Content Ratings and Credibility Indicators | 2005 | 29 | 2 |
| 20 | Chungsuk Ryu, Yong Jin Kim, Abhijit Chaudhury, H. Raghav Rao | Knowledge Acquisition via Three Learning Processes in Enterprise Information Portals: Learning-by-Investment, Learning-by-Doing, and Learning-from-Others | 2005 | 29 | 2 |
| 21 | Huseyin Tanriverdi | Information Technology Relatedness, Knowledge Management Capability, and Performance of Multibusiness Firms | 2005 | 29 | 2 |
| 22 | Andrew H. Van de Ven | Running in Packs to Develop Knowledge-Intensive Technologies | 2005 | 29 | 2 |
| 23 | Vicky Arnold, Nicole Clark, Philip A. Collier, Stewart A. Leech, and Steve G. Sutton | The Differential Use and Effect of Knowledge-Based System Explanations in Novice and Expert Judgement Decisions | 2006 | 30 | 1 |
| 24 | Anne P. Massey and Mitzi M. Montoya-Weiss | Unraveling the Temporal Fabric of Knowledge Conversion: A Model of Media Selection and Use | 2006 | 30 | 1 |
| 25 | Victoria L. Mitchell | Knowledge Integration and Information Technology Project Performance | 2006 | 30 | 4 |
| 26 | Satish K. Puri | Integrating Scientific with Indigenous Knowledge: Constructing Knowledge Alliances for Land Management in India | 2007 | 31 | 2 |
| 27 | Prasert Kanawattanachai and Youngjin Yoo | The Impact of Knowledge Coordination on Virtual Team Performance Over Time | 2007 | 31 | 4 |
| 28 | Hoon S. Cha, David E. Pingry, and Matt E. Thatcher | Managing the Knowledge Supply Chain: An Organizational Learning Model of Information Technology Offshore Outsourcing | 2008 | 32 | 2 |
| 29 | Paul M. Leonardi and Diane E. Bailey | Transformational Technologies and the Creation of New Work Practices: Making Implicit Knowledge Explicit in Task-Based Offshoring | 2008 | 32 | 2 |
| 30 | Sue Young Choi, Heeseok Lee, and Youngjin Yoo | The Impact of Information Technology and Transactive Memory Systems on Knowledge Sharing, Application, and Team Performance: A Field Study | 2010 | 34 | 4 |
| 31 | Palash Bera, Andrew Burton-Jones, and Yair Wand | Guidelines for Designing Visual Ontologies to Support Knowledge Identification | 2011 | 35 | 4 |
| 32 | Jessica Luo Carlo, Kalle Lyytinen, and Gregory M. Rose | A Knowledge-Based Model of Radical Innovation in Small Software Firms | 2012 | 36 | 3 |
| 33 | Yinglei Wang, Darren B. Meister, and Peter H. Gray | Social Influence and Knowledge Management Systems Use: Evidence from Panel Data | 2013 | 37 | 1 |
| 34 | Anne Majchrzak, Christian Wagner, and Dave Yates | The Impact of Shaping on Knowledge Reuse for Organizational Improvement with Wikis | 2013 | 37 | 2 |
| 35 | Julia Kotlarsky, Harry Scarbrough, and Ilan Oshri | Coordinating Expertise Across Knowledge Boundaries in Offshore-Outsourcing Projects: The Role of Codification | 2014 | 38 | 2 |
| 36 | Roman Beck, Immanuel Pahlke, and Christoph Seebach | Knowledge Exchange and Symbolic Action in Social Media-Enabled Electronic Networks of Practice: A Multilevel Perspective on Knowledge Seekers and Contributors | 2014 | 38 | 4 |
| 37 | Kishen Iyengar, Jeffrey R. Sweeney, and Ramiro Montealegre | Information Technology Use as a Learning Mechanism: The Impact of IT Use on Knowledge Transfer Effectiveness, Absorptive Capacity, and Franchisee Performance | 2015 | 39 | 3 |
| 38 | Paul M. Leonardi | Ambient Awareness and Knowledge Acquisition: Using Social Media to Learn "Who Knows What" and "Who Knows Whom" | 2015 | 39 | 4 |
| 39 | Seung Hyun Kim, Tridas Mukhopadhyay, and Robert E. Kraut | When Does Repository KMS Use Lift Performance? The Role of Alternative Knowledge Sources and Task Environments | 2016 | 40 | 1 |
| 40 | Peng Huang and Zhongju Zhang | Participation in Open Knowledge Communities and Job-Hopping: Evidence from Enterprise Software | 2016 | 40 | 3 |
| 41 | Harris Kyriakou, Jeffrey V. Nickerson, and Gaurav Sabnis | Knowledge Reuse for Customization: Metamodels in an Open Design Community for 3D Printing | 2017 | 41 | 1 |
| 42 | Konstantinos Trantopoulos, Georg von Krogh, Martin W. Wallin, and Martin Woerter | External Knowledge and Information Technology: Implications for Process Innovation Performance | 2017 | 41 | 1 |
| 43 | Yingda Lu, Param Vir Singh, and Baohong Sun | Is a Core-Periphery Network Good for Knowledge Sharing? A Structural Model of Endogenous Network Formation on a Crowdsourced Customer Support Forum | 2017 | 41 | 2 |
| 44 | Xiaojun Zhang | Knowledge Management System Use and Job Performance: A Multilevel Contingency Model | 2017 | 41 | 3 |

All errors remain with the curation team.